\documentclass[aps,prc,twocolumn,10pt,superscriptaddress,floatfix,showpacs]{revtex4-1}
\usepackage{graphicx}
\usepackage{amsmath}
\usepackage{tabularx}

\begin{document}
\title{The Proton Radius from Electron Scattering Data}

\author{Douglas W. Higinbotham}
\affiliation{Jefferson Lab, 12000 Jefferson Avenue, Newport News, Virginia 23606 USA}

\author{Al Amin Kabir}
\affiliation{Kent State University, 800 E. Summit St., Kent State, Ohio, 44240 USA}

\author{Vincent Lin}
\affiliation{Jefferson Lab, 12000 Jefferson Avenue, Newport News, Virginia 23606 USA}
\affiliation{Western Branch High School, 1968 Bruin Place, Chesapeake, VA 23321 USA}

\author{David Meekins}
\affiliation{Jefferson Lab, 12000 Jefferson Avenue, Newport News, Virginia 23606 USA}

\author{Blaine Norum}
\affiliation{University of Virginia, Charlottesville, Virginia 22904 USA}

\author{Brad Sawatzky}
\affiliation{Jefferson Lab, 12000 Jefferson Avenue, Newport News, Virginia 23606 USA}

\begin{abstract}
\begin{description}
\item[Background] The proton charge radius extracted from recent 
muonic hydrogen Lamb shift measurements is significantly smaller 
than that extracted from atomic hydrogen and electron scattering 
measurements.   The discrepancy has become known as the proton 
radius puzzle.
\item[Purpose] In an attempt to understand the discrepancy, we review 
high-precision electron scattering results from Mainz, Jefferson Lab, 
Saskatoon and Stanford.
\item[Methods] We make use of stepwise regression techniques using 
the $F$-test as well as the Akaike information criterion to systematically 
determine the predictive variables to use for a given set and range 
of electron scattering data as well as to provide multivariate error 
estimates.
\item[Results] Starting with the precision, low four-momentum transfer 
($Q^2$) data from Mainz (1980) and Saskatoon (1974), we find that 
a stepwise regression of the Maclaurin series using the $F$-test as well as 
the Akaike information criterion justify using a linear extrapolation 
which yields a value for the proton radius that is consistent with 
the result obtained from muonic hydrogen measurements.
Applying the same Maclaurin series and statistical criteria 
to the 2014 Rosenbluth results on $G_E$ from Mainz, we again find
that the stepwise regression tends to favor a radius consistent 
with the muonic hydrogen radius but produces results that are extremely
sensitive to the range of data included in the fit.  Making use of 
the high-$Q^2$ data on $G_E$ to select functions which extrapolate 
to high $Q^2$, we find that a Pad\'e ($N=M=1$) statistical model 
works remarkably well, as does a dipole function with a 0.84~fm radius, 
$G_E(Q^2) = ( 1 + Q^2/0.66\,\mathrm{GeV}^2)^{-2}$.
\item[Conclusions] Rigorous applications of stepwise regression techniques 
and multivariate error estimates result in the extraction of a proton 
charge radius that is consistent with the muonic hydrogen result 
of 0.84~fm; either from linear extrapolation of the extreme low-$Q^2$ data 
or by use of the Pad\'e approximant for extrapolation using a larger range
of data.   Thus, based on a purely statistical analysis of electron 
scattering data, we conclude that the electron scattering result 
and the muonic hydrogen result are consistent.
It is the atomic hydrogen results that are the outliers.
\end{description}
\end{abstract}
\maketitle

The visible universe is primarily comprised of protons, yet many 
of this particle's properties are not yet well understood, including 
its charge radius.  In recent muonic hydrogen Lamb shift experiments, 
it was determined that the proton's charge radius
is 0.8409(4)~fm \cite{Pohl:2010,Antognini:2012}.   
This result is in significant disagreement with the 2010 CODATA 
recommended value of 0.878(5)~fm based on spectroscopic data 
and elastic electron scattering fits \cite{Mohr:2012}.  
The muonic result is also in disagreement with the most recent value of
0.879$(5)_\mathrm{stat}(4)_\mathrm{sys}(2)_\mathrm{model}(4)_\mathrm{group}$~fm
obtained from electron scattering \cite{Bernauer:2013tpr}.  
This disagreement has become known as 
the proton radius puzzle~\cite{Pohl:2013yb,Carlson:2015jba}.

%
%
%

In contrast to other groups that have focused on re-analysis of recent
data \cite{Griffioen:2015hta, Horbatsch:2015qda, Lee:2015jqa, Lorenz:2014yda}, 
our study began by reviewing the high-precision experiments from Mainz
\cite{Simon:1980hu} and Saskatoon \cite{Murphy:1974zz}, referred to herein 
as Mainz80 and Saskatoon74.  These low-$Q^2$ ($0.14$--$1.4$~fm$^{-2}$) 
electron scattering measurements were made using hydrogen gas targets.
The Saskatoon74 measurements involved detecting the recoiling protons,
while the Mainz80 measurements involved detecting the scattered electrons.
These were both extremely high precision experiments, with great care 
taken to control point-to-point systematic uncertainties.
Prior to these measurements, the value of the proton radius 
was generally believed to be 0.81(1)~fm \cite{Hand:1963zz}, 
while after Mainz80 data the accepted value from electron scattering 
began approaching the current CODATA value.

As noted by Hand {\it{et al.}}~\cite{Hand:1963zz}, one clear advantage 
of using extremely low-$Q^2$ data for analysis of the proton's charge radius 
is that experimental cross sections are dominated by the electric
(charge) form factor, $G_E$, making the results relatively insensitive 
to the magnetic form factor.  Also, since the low-$Q^2$ data are fit 
such that $G_E(0) \equiv 1$, corrections which shift all points 
at once are automatically taken into account. 
Of course, the major disadvantage of using just the lowest $Q^2$ is the
relatively limited amount of high precision data; though, paradoxically, 
not all {\it{global}} fits include the Saskatoon74 data 
(e.g.~\cite{Arrington:2003qk, Arrington:2007ux, Lee:2015jqa}).

In principle, the charge radii of nuclei can be determined from elastic 
electron scattering data by fitting the extracted charge form factors 
and using the Fourier transform to extract the corresponding 
charge distribution and rms radius \cite{DeJager:1987qc,Cardman:1980}.
For heavy nuclei, the locations of diffraction minima of the form factor
also provide a significant constraint on the radius.
For the proton, however, no such diffraction minimum
has been found and, due to the light masses of the quarks within the proton, 
relativistic corrections preclude simply using the Fourier transform
to extract its charge density \cite{Miller:2007uy}.

Instead, the proton's charge radius, $r_p$, can be extracted 
by determining the slope of the electric form factor as $Q^2$ 
approaches zero.  The relation between the slope and the radius 
can be derived in the Born approximation by integrating out 
the angular dependence of the form factor, resulting in the series 
$$
G_E(Q^2) 
  =  1 
  +  \sum_{n\ge 1} \frac{(-1)^n}{(2n+1)!}  
     \left\langle r^{2n} \right\rangle \, Q^{2n} \>.
$$
Hence, $r_p$ can be determined from
$$
  r_p \equiv \sqrt{ \langle r^2 \rangle} 
  = \left( -6  \left. \frac{\mathrm{d} G_E(Q^2)}{\mathrm{d}Q^2} 
    \right|_{Q^{2}=0} \right)^{1/2} \>.
$$

Of course, electron scattering cannot reach the exact $Q^2 = 0$ limit.
Thus, we must extrapolate from the lowest $Q^2$ data available to $Q^2=0$.
In order to more accurately determine a slope at the origin, a range of data 
over the smallest experimentally accessible $Q^2$ values is favorable.
There are many methods which may be employed for extrapolation; 
one such method involves fitting a model to available data over 
a selected range, typically nearest to the interval over which 
the extrapolation is to be applied.  The 1963 fit to 
(and extrapolation of) the $G_E(Q^2)$ data \cite{Hand:1963zz} 
determined the proton radius to be 0.805~fm by using the expansion
\begin{equation}
\label{maclaurin}
  f(Q^2) 
  = n_0  G_E(Q^2) 
  \approx n_0 \biggl( 1 + \sum_{i=1}^m a_i Q^{2i} \biggr) \>.
\end{equation}
Here $n_0$ is a normalization factor in recognition that certain 
experimental parameters, such as beam current and target thickness, 
are not known exactly and variations in them will affect 
all elements of a data set the same way. 

While extrapolations to regions beyond available data can be misleading,
requiring $G_E(Q^2=0)=1$ alleviates some of the uncertainty.  In general,
extrapolations using linear regression with the points nearest to the
extrapolation region can be employed with reasonable success.
When using polynomial extrapolations, both the interval over which
the model is fit and the degree of the polynomial
must be carefully chosen to avoid pathological behaviors.
In particular, while adding higher order terms will improve the
$\chi^2$, they can also magnify small statistical or systematic deviations,
as can be seen in the recent Monte Carlo study~\cite{Kraus:2014qua}.
Thus great care must be taken to use only statistically justified terms
when fitting a given data set.

%
%

To mitigate the effects of using an inappropriate number of variants,
standard statistical methods were used to determine the 
degree of polynomial to fit a given data set spanning a given interval. 
Taking the Mainz80 and Saskatoon74 data in their overlap region, 
$0.14 \leq Q^2 \leq 0.8$~fm$^{-2}$, $F$-tests were used to compare 
first- and second-order polynomial fits.  These tests indicated 
that the higher order term was not statistically significant in this region; 
thus, the $Q^4$ term was rejected (see Appendix~\ref{polyapp}). 
We also made use of a full stepwise regression analysis using 
the Akaike information criterion code and obtained the same result 
(see \cite{Higinbotham:2016}).

Accordingly, linear extrapolation functions were chosen as 
the statistical model of the combined Saskatoon74 and Mainz80 data 
with cut-offs up to $Q^2 \leq 0.8$~fm$^{-2}$ as shown in Table~\ref{taylort}.
As this is a multivariate fit, where both the slope and intercept 
are simultaneously needed, we use the two-parameter $\Delta\chi^2$ 
of 2.30 \cite{James:2006} to estimate the 68.3\% probability content. 
The fit was done using MINUIT's order-independent $\chi^2$ minimization with
$$
\chi^{2} 
  = \sum\nolimits_{i} \left( f(Q^2)_i - \mathrm{data}(Q^2)_i \right)^2  
    / \left( \sigma(Q^2)_i \right)^2 \>.
$$
Thus, we do not attempt to shift the normalization of the data 
as was done in some works, but instead have added flexibility 
to the model via an $n_0$ term (see Eq.~\ref{maclaurin}) to allow for 
possible normalization differences. 

\begin{table}
\caption{Linear extrapolations
to $Q^2=0$ (Eq.~\ref{maclaurin} with $m=1$)
of the Saskatoon74 and Mainz80 data
for various $Q^2$ cut-offs (upper limits).
These  linear fits/extrapolations
are used solely to find the slope and intercept at $Q^2=0$
and have no meaning beyond the cut-off.}
\label{taylort}
\begin{ruledtabular}
\begin{tabularx}{1.0\textwidth}{cclrcl}
$Q_\mathrm{max}^2\,[\mathrm{fm}^{-2}]$   &  $n_0$     & $a_1\,[\mathrm{fm}^2]$       & ${\chi}^2$ &   ${\chi}_\mathrm{red}^2$   & $r_p\,[\mathrm{fm}]$  \\ 
\hline
0.4       & 1.000(4) & $-0.111(14)$ &  6.8      & 0.684   & 0.816(51) \\
0.5       & 1.002(3) & $-0.110(10)$ &  8.9      & 0.636   & 0.840(35) \\
0.6       & 1.001(3) & $-0.118(7)$  &  9.3      & 0.547   & 0.840(27) \\
0.7       & 1.002(2) & $-0.120(4)$  & 10.7      & 0.534   & 0.851(16) \\
0.8       & 1.002(2) & $-0.119(4)$  & 13.7      & 0.623   & 0.844(14) \\
\end{tabularx}
\end{ruledtabular}
\end{table}

It is important to note that all statistical quantities 
such as standard errors, confidence intervals, etc., while rigorously 
determined in the region where existing data are used in the fit, 
are not as accurate over the extrapolated region.  But given the small range 
of the extrapolation, we nonetheless apply them to estimate the uncertainty
of the charge radius.  The results obtained with the linear extrapolations 
are all within one sigma (68.3\% confidence interval) of the radius 
extracted from the muonic Lamb shift data.   These results are also 
in statistical agreement with the $r_p = 0.840(16)$~fm found by 
Griffioen, Carlson, and Maddox~\cite{Griffioen:2015hta} with 
their power series (Eq.~\ref{maclaurin} with $m=1,2$) and continued-fraction 
fits of the new high-precision low-$Q^2$ Mainz14 data~\cite{Bernauer:2013tpr}.  

These results seem to systematically disagree with the Mainz14 reported 
radius of 0.88~fm where much higher order functions (9th and 10th 
degree polynomial and spline functions \cite{Bernauer:2010wm}) 
and large $Q^2$ ranges were used to extract the proton radius.
Now, while the high-order functions can be used for precisely 
interpolating the data those same functions are not typically used 
for accurate extrapolations.  This same tension between low-$Q^2$ 
cut-off vs.~high-$Q^2$ cut-off extrapolations 
can be explicitly seen in the recent global fits of Lee, Hill 
and Arrington \cite{Lee:2015jqa} where they systematically get 
a smaller proton radius when they use only low-$Q^2$ data 
and a much larger charge radius as they include high-$Q^2$ data.

To investigate how the next degree of the fitting polynomial affects 
the extracted charge radius, we employed a three-parameter fit 
(Eq.~\ref{maclaurin} with $m=2$) encompassing the full range
of the Mainz80 ($0.14-1.4$~fm$^{-2}$) and Saskatoon74 ($0.15-0.8$~fm$^{-2}$) 
data.  With the full range of these data, one is now statistically justified 
in including the $Q^4$ term as can be seen in the ANOVA table 
in Appendix~\ref{ANOVA-tables}.  For the three-parameter fits, 
the values, $X$, and the covariance matrix, $\Sigma$, are given by
$$
    X =
\begin{pmatrix}
 n_0 \\
 a_1 \\
 a_2 \\
\end{pmatrix},~
 \Sigma = 
\begin{pmatrix}
\sigma_{0}^2                     & \rho_{01} \sigma_{0} \sigma_{1} & \rho_{02} \sigma_{0} \sigma_{2} \\
\rho_{10} \sigma_{0} \sigma_{1} & \sigma_{1}^2                     & \rho_{12} \sigma_{1} \sigma_{2} \\
\rho_{20} \sigma_{0} \sigma_{2} & \rho_{21} \sigma_{1} \sigma_{2} & \sigma_{2}^2 \\
\end{pmatrix},
$$
where $\sigma_i$ are the uncertainties and $\rho_{ij}$ are 
the correlation coefficients.  As we are not using orthogonal polynomials, 
there will be significant correlations between the fit parameters.
An example of the effect of these correlations can be seen by simply 
shifting the highest $Q^2$ point in the combined Mainz80 and Saskatoon74 
data set by one standard deviation and refitting the data.  This one small 
change has the effect of systematically shifting all three fit parameters 
and changing the extracted proton radius by 0.010~fm.   
This is a substantial amount, considering the difference between 
the CODATA and the muonic hydrogen values is only 0.037~fm.

The confidence region for the parameter set of a multivariate fit 
can be defined by  
$$
\chi^2 \leq \chi^2_{\mathrm{min}} + \Delta\chi^2,
$$
where $\chi^2_{\mathrm{min}}$ is the $\chi^2$ found when fitting 
the data and $\Delta\chi^2$ (also called $K_{\beta}^2$~\cite{James:2006}) 
defines the confidence region.  By doing this, we are generating 
a covariance matrix normalized to our desired confidence level 
(see Appendix~\ref{polyapp} for details). Thus, in order to have 
a 68.3\% probability content for a three-parameter multivariate fit, 
one should use $\Delta\chi^2$= 3.53 (see Table 38.2 of the Review 
of Particle Properties~\cite{PDG:2014}).  Using this confidence region, 
a fit of the combined Mainz80 and Saskatoon74 data yields
\begin{equation}
\label{eqn:covar}
    X =
\begin{pmatrix}
  1.003 \\
 -0.127 \\
  0.011 \\
\end{pmatrix},~
\Sigma = 
\begin{pmatrix}
 1.26&
-3.66 &
 2.46 \\
-3.66 &
 12.6 &
-9.26 \\
 2.46 &
-9.26 &
 7.44 \\
\end{pmatrix}\times 10^{-5}.
\end{equation}
Thus one finds $n_0=1.003(4)$, $a_1=-0.127(11)$ and $a_2=0.011(8)$,
resulting in a radius of $r_p=0.873(39)$~fm.  This is within one sigma 
of the muonic result where we have again assumed that uncertainties 
determined for the fitted region are valid when extrapolated to $Q^2=0$.

\begin{figure}[hbtp]
\includegraphics[width=\linewidth]{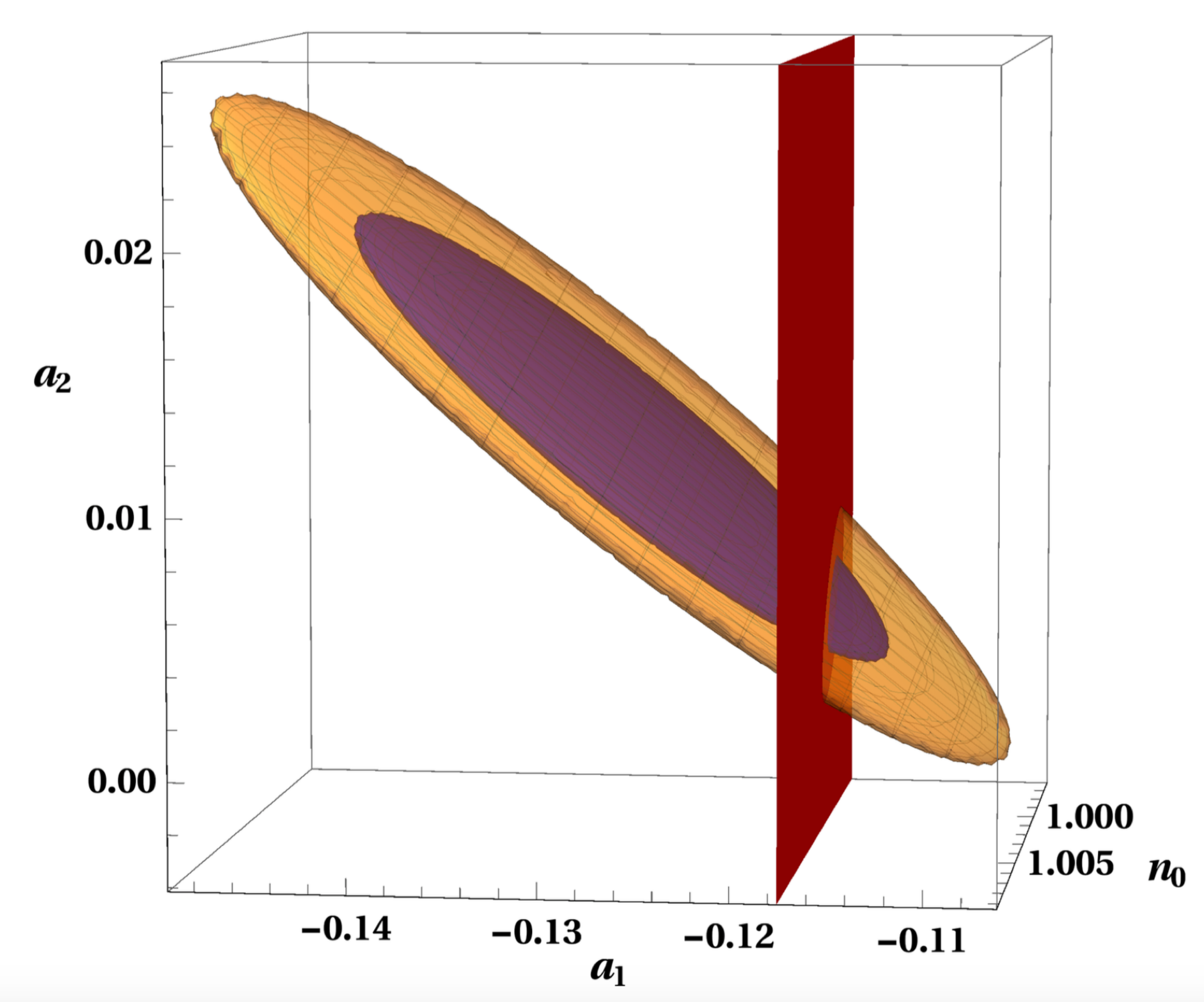}
\caption{The 68\% (inner) and 95\% (outer) confidence ellipsoids 
associated with the covariance matrix from the three-parameter fit 
(Eq.~\ref{eqn:covar}) of the Mainz80 and Saskatoon74 data.
The plane representing the muonic Lamb shift result of $r_p = 0.84$~fm
is shown at its corresponding $a_1$ value of $-0.1176$~fm$^2$ 
and is clearly not ruled out by this fit.}
\label{pretty}
\end{figure}

To visualize the parameter correlations, we generated both the 68\% and 95\% 
confidence ellipsoids as well as the plane indicating the $a_1$ given by 
the muonic hydrogen Lamb shift result (Fig.~\ref{pretty}).   
The intersection of these regions shows that the muonic hydrogen result 
is within even the tighter confidence ellipsoid.

We note that the statistical models,
which are being used to extrapolate the slope of $G_E$ to $Q^2=0$, 
are only valid near where they are constrained by data and are not intended for
extrapolation far beyond the data that is being fit.   Even so, it is important
to research whether the function being used can reasonably be expected to
extrapolate reliably.   In mathematics, there is well known tension between
high order functions which precisely interpolate a set of data and lower order 
functions which can accurately extrapolate.

\begin{figure}[hbtp]
  \centering
  \includegraphics[width=0.96\linewidth]{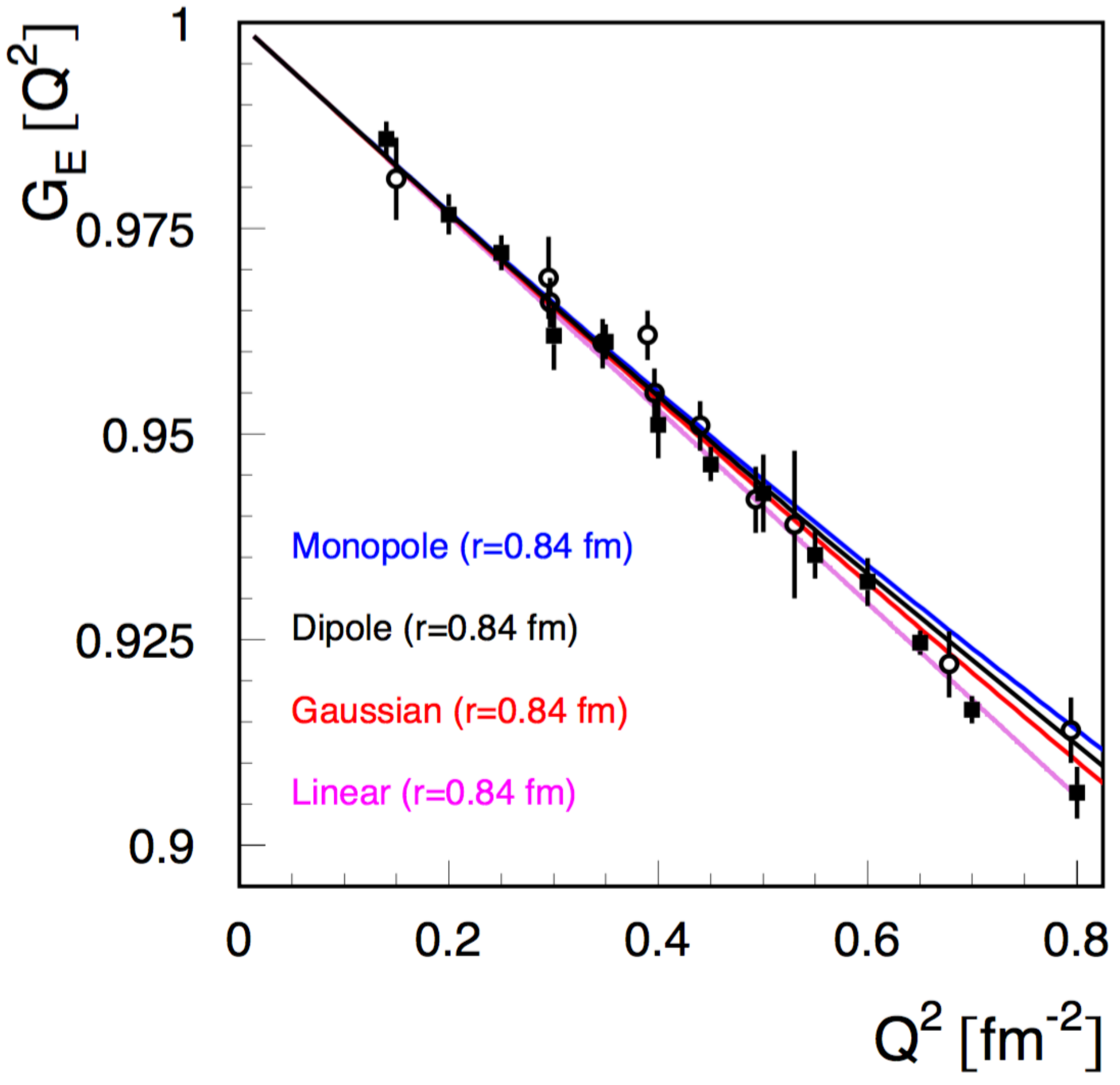}
  \includegraphics[width=0.96\linewidth]{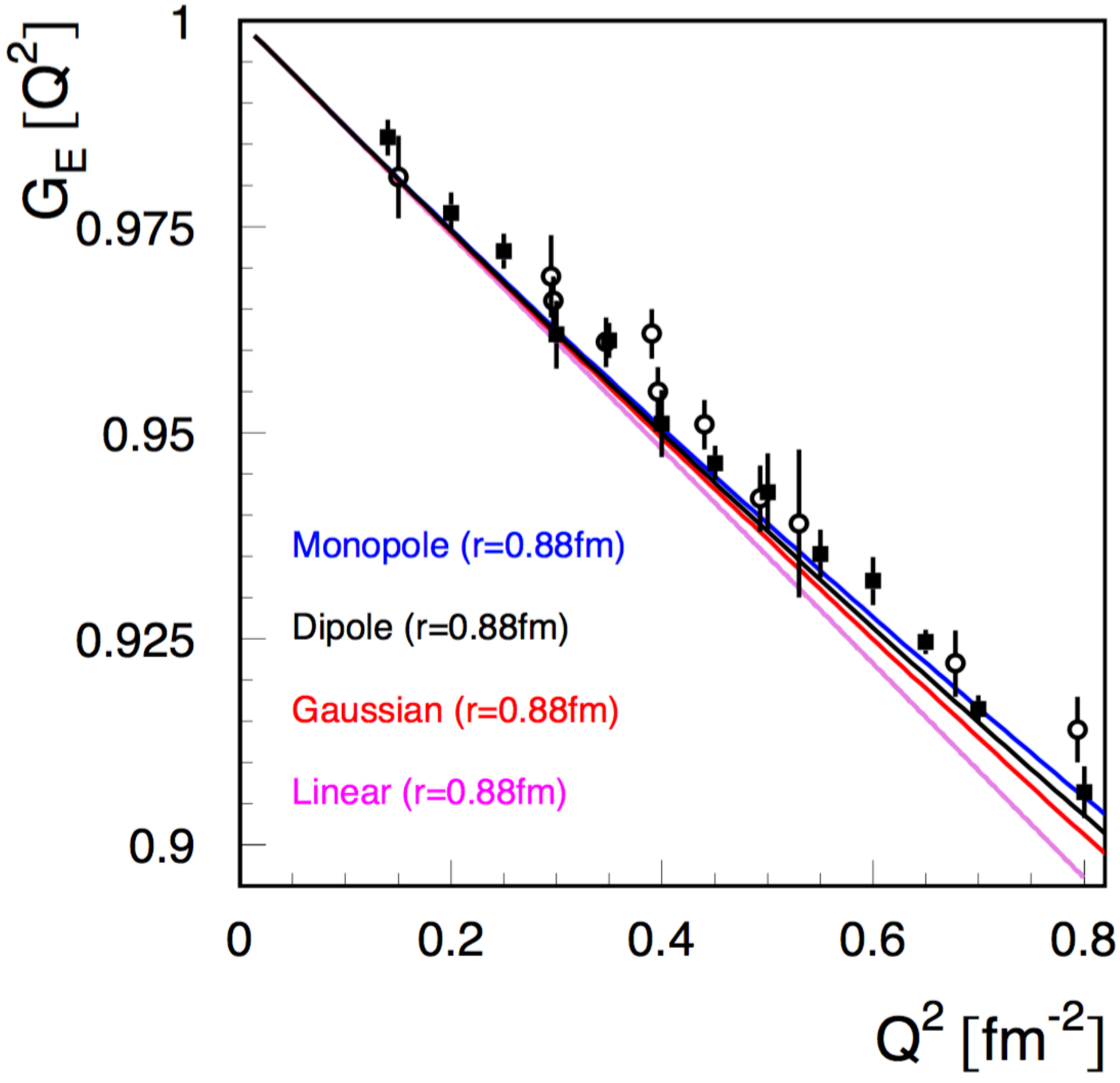}
  \caption{Shown are monopole, dipole, Gaussian, and the linear 
extrapolations with the radius set to the muonic Lamb shift result 
of 0.84~fm (top panel) and 0.88~fm (bottom panel) for $Q^2$ 
up to 0.8~fm$^{-2}$, along with the high-precision, 
low-$Q^2$ data from Saskatoon74, open circles, and Mainz80, solid squares.
At very low $Q^2$ these functions overlap as we know they should 
from their power-series expansions.}
  \label{TheZoom}
\end{figure}

Since the linear extrapolations generally agree with the muonic 
hydrogen Lamb shift result, we tried simply using the muonic hydrogen 
radius, $0.84$~fm, to set the slope parameter, $a_1 = -0.1176$~fm$^2$, 
in the linear, monopole, dipole and Gaussian functions.  The results 
are shown in Fig.~\ref{TheZoom} (top panel) with the published $G_E(Q^2)$ 
results, i.e., without any renormalization, for $0< Q^2 < 0.8$~fm$^{-2}$.   
The linear extrapolation shown in this plot is indistinguishable 
from the $a_1$ of $-0.1163$~fm$^2$ (i.e., $r_p = 0.835(3)$~fm) 
found in the low-$Q^2$ re-analysis of the Mainz14 data by Griffioen, 
Carlson and Maddox~\cite{Griffioen:2015hta}.  Interestingly, these results 
are also in agreement with the inverse-polynomial fits of Arrington 2004
\cite{Arrington:2003qk} where the slopes at $Q^2=0$ yield radii between 
0.835 and 0.856~fm.   Many other fits, such as found in Mainz14
\cite{Bernauer:2013tpr}, yield larger radii.   The latter fits include 
far more free parameters, involve data from a wider range of $Q^2$,
and do not go through the published low-$Q^2$ values of $G_E$.
Figure~\ref{TheZoom} (bottom panel) shows the same as the panel above,
but with $r_p = 0.88$~fm.   It is important to realize that in a global
$\chi^2$ minimization with floating normalizations, the lowest-$Q^2$ 
data set can easily be shifted by the relatively large amount of higher-$Q^2$ 
data.

To further investigate the source of this discrepancy, we next
studied the Mainz $G_E(Q^2)$ Rosenbluth data~\cite{Bernauer:2013tpr}.   
For these fits, we allow for a $0.2\%$ systematic uncertainty to 
the statistical uncertainty in line with other very high precision 
measurements that were not statistics-limited~\cite{Cardman:1980}.

Using the simplest function that has not been ruled out for this
range in $Q^2$, 
the dipole function, we first fit this data set alone and find 
\begin{equation*}
X =
\begin{pmatrix}
 0.991  \\
-0.1140 \\
\end{pmatrix},~
\Sigma=
\begin{pmatrix}
18.16 & 5.449 \\ 
5.449 & 2.260 \\
\end{pmatrix}\times 10^{-7},
\end{equation*}
with a reduced $\chi^2$ of 0.725.
This parameter set yields $r_p =0.827(2)$~fm and again
favors the muonic hydrogen result.   We next perform Maclaurin 
series fits, Eq.~\ref{maclaurin}, using the $F$-test
to determine the optimal order of the fits (see supplemental material
and Fig.~\ref{TheRosen}) and find $r_p = 0.823(2)$~fm and 
$r_p = 0.857(2)$~fm from the 5- and 6-parameter fits, respectively.
Finally, we fit using a rational ($N=M=1$ Pad\'{e}) approximant
$$
f(x) = n_0 \frac{1 + a_1 x}{1 + b_1 x} \>,
$$
a well behaved  function often used for extrapolations \cite{sirca:2012} 
and again use an $F$-test to determine the order of the fit.
Here the radius at $Q^2=0$ is given by $\sqrt{6(b_1-a_1)}$.  We fit
these 77 points using $\Delta\chi^2 = 3.51$ to find 
\begin{equation*}
X =
\begin{pmatrix}
 0.994  \\
-0.0193 \\
 0.0980 \\
\end{pmatrix},~
\Sigma=
\begin{pmatrix}
5.74 & 1.41  & 4.58 \\ 
1.41 & 0.825 & 1.97 \\
4.58 & 1.97  & 5.29 \\
\end{pmatrix}\times 10^{-6},
\end{equation*}
with a reduced $\chi^2$ of 0.624 and a radius of $r_p = 0.839(9)$~fm.

We also tried fixing the radius to 0.84~fm and 0.88~fm 
(i.e.~$a_1$ = 0.1176~fm$^2$ and 0.1292~fm$^2$, respectively) 
and performing a five-parameter Maclaurin fits of the Mainz14 Rosenbluth $G_E$ data:
see Table~\ref{RosenFixR}.  Interestingly, even here the $\chi^2$ is significantly 
better for the smaller radius.  

\begin{table*}
\caption{Repeating the Maclaurin $j=6$ fits, but with the $a_1$ term 
fixed to the atomic hydrogen and muonic hydrogen values of the proton 
radius, 0.84~fm and 0.88~fm.}
\label{RosenFixR}
\begin{ruledtabular}
\begin{tabular}{cccccccc}
Fixed Radius & $\chi^2$  & $\chi^2/{\nu}$ & $n_0$    & $a_2$                  & $a_3$                   & $a_4$                  & $a_5$                 \\
\hline
0.84~fm      & 56.34     & 0.783          & 0.994(1) & $1.12(1)\cdot 10^{-2}$ & $-0.93(2)\cdot 10^{-3}$ & $5.0(1) \cdot 10^{-5}$ & $ 1.20(5)\cdot 10^{-6}$ \\ 
0.88~fm      & 142.1     &  1.97          & 1.003(1) & $1.62(1)\cdot 10^{-2}$ & $-1.78(1)\cdot 10^{-3}$ & $1.14(1)\cdot 10^{-4}$ & $-2.90(7)\cdot 10^{-6}$ \\
\end{tabular}
\end{ruledtabular}
\end{table*}

Recent asymmetry data have shown that $G_E$ stays
positive beyond 220~fm$^{-2}$~\cite{Puckett:2010ac}, 
so we next decided to compare our simple form factor functions 
to the highest measured values of $G_E$.  We repeated our two-parameter 
dipole function
fits of the low-$Q^2$ Mainz80 and Saskatoon74 data, but now included 
the Stanford94 \cite{Andivahis:1994rq} and Jefferson~Lab04
\cite{Christy:2004rc} Rosenbluth results.  By including these data, 
the monopole and Gaussian functions can be rejected with a high degree 
of certainty due to their reduced $\chi^2$ of 34 and 25, respectively~%
\footnote{While a large reduced $\chi^2$ can be used to reject 
a model with a high degree of certainty, $\chi^2$ alone should 
not be used for choosing between models that have not been rejected.}.
On the other hand, the two-parameter dipole form factor fit of these 
four independent data sets, two at extremely low $Q^2$ and two 
at extremely high $Q^2$, yields
\begin{equation*}
X =
\begin{pmatrix}
 0.9995 \\
-0.1201
\end{pmatrix},~
\Sigma =
\begin{pmatrix}
   2.443 & -3.430 \\
  -3.430 &  6.773 \\
\end{pmatrix}\times 10^{-6},
\end{equation*}
with a reduced ${\chi}^2$ of 1.25 and a normalization of nearly one.
If the $a_1$ term of this fit is interpreted in terms of a charge radius,
one finds $r_p$ = 0.848(9)fm.  This two-parameter fit, perfectly describing 
and  bridging the lowest and highest $G_E(Q^2)$ results, is again consistent 
with the muonic hydrogen Lamb shift radius.  Repeating as a one-parameter fit,
i.e.~$n_0 = 1$ from the low-$Q^2$ fits above, gives $a_1=-0.1188(14)$~fm$^2$ 
for a radius of 0.844(5)~fm and reduced $\chi^2$ of 1.23.  We note that 
the model-dependent part of the two-photon exchange correction can reduce
the high-$Q^2$ $G_E$ values, but we also know that $G_E$ remains 
positive \cite{Puckett:2010ac}.  Thus, the absolute range of high-$Q^2$
$G_E$ values is actually quite small and within the included systematic 
uncertainty.

%
%
\begin{figure}[hbtp]
\includegraphics[width=\linewidth]{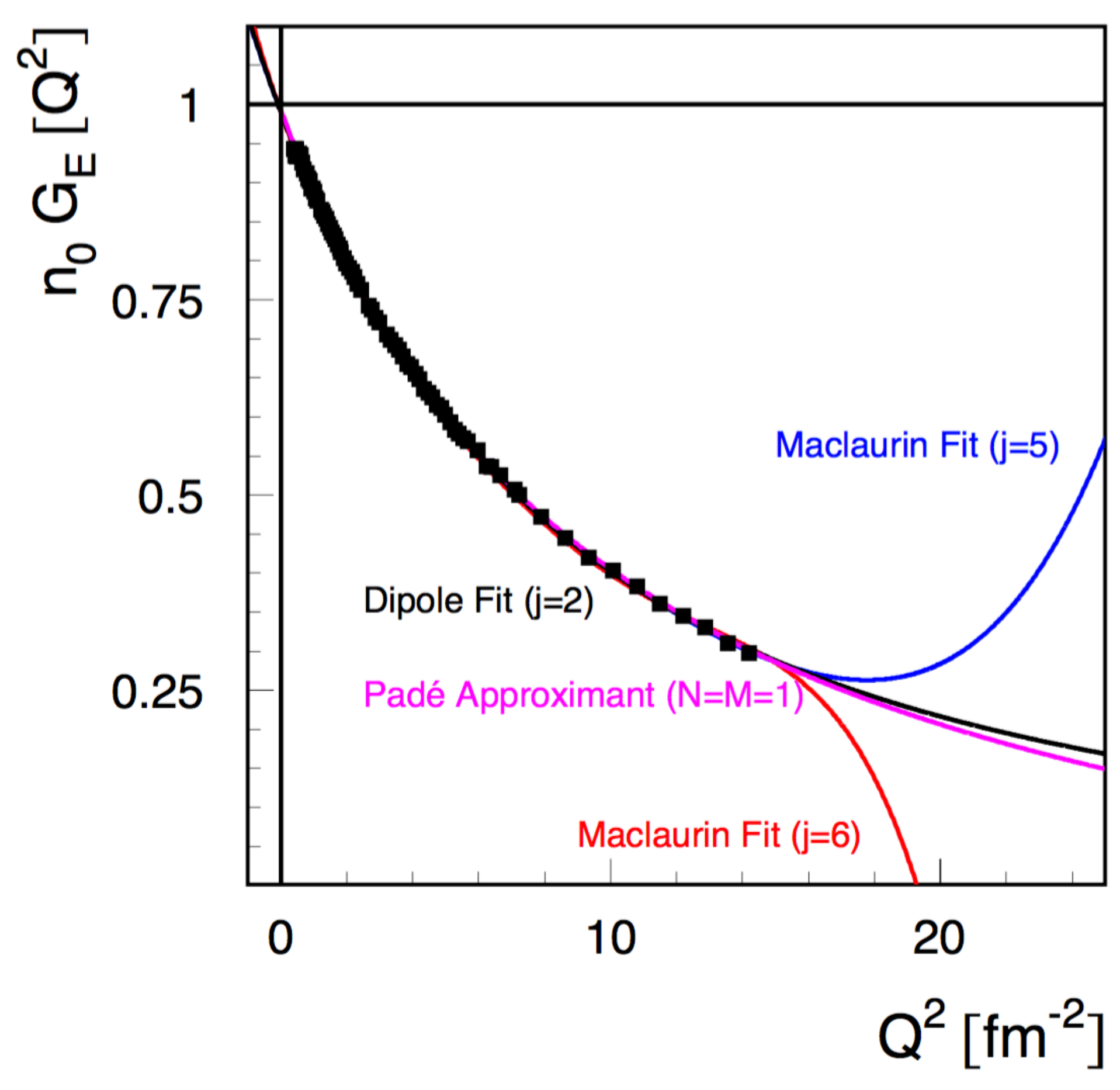}
\caption{Shown are various fits to Mainz14 Rosenbluth $G_E$ data.   
One can see that Maclaurin series fits quickly diverge when not 
constrained by the data.  The dipole (a special case of a rational function) 
and the $N=M=1$ Pad\'e approximant not only extrapolate to high $Q^2$ 
reasonably well, but unlike the Maclaurin series produces nearly 
the same $Q^2=0$ slope and intercept for different cut-offs.}
\label{TheRosen}
\end{figure}

%
%

\begin{figure}[hbtp]
  \centering
  \includegraphics[width=\linewidth]{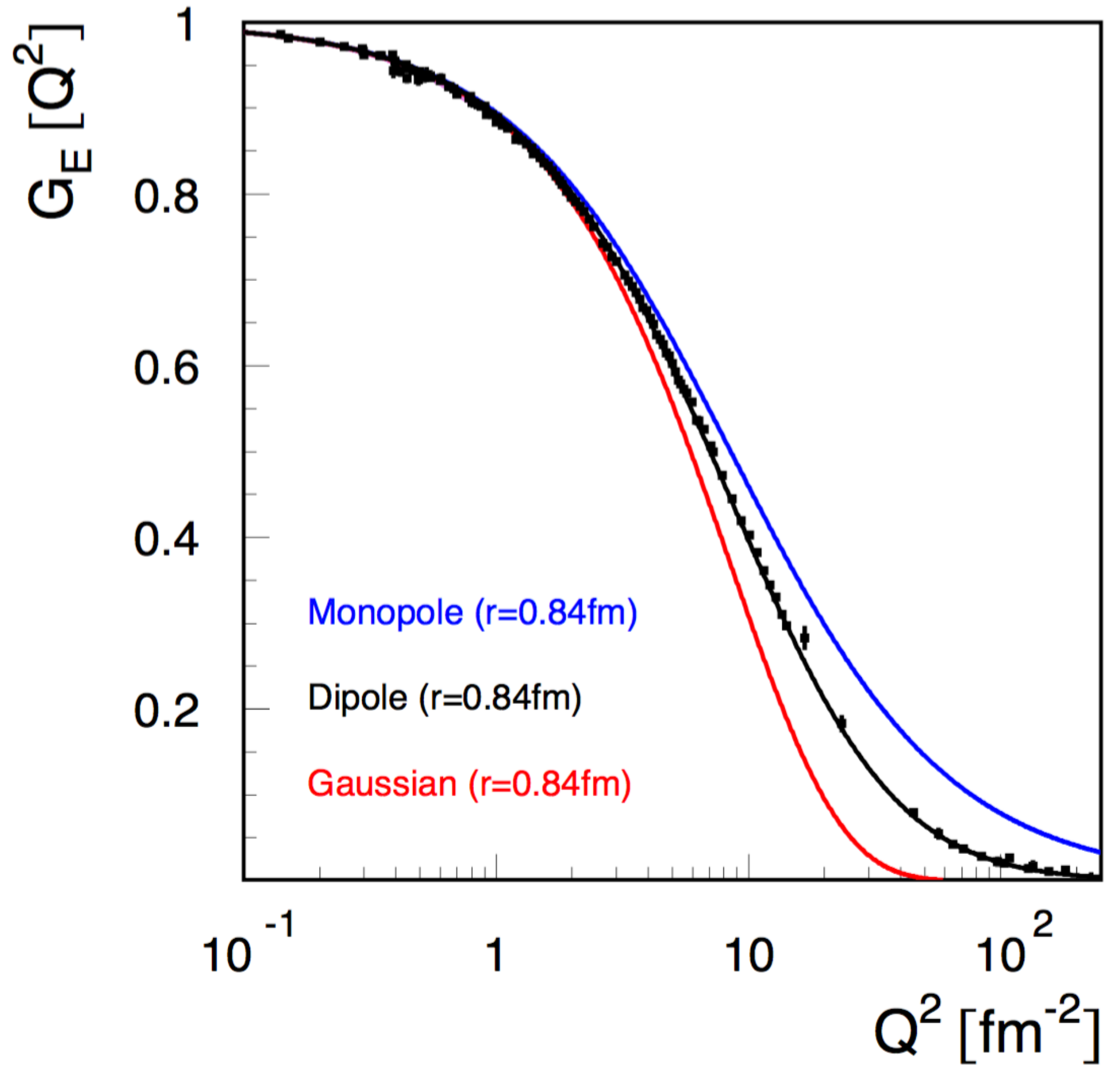}
  \caption{Shown are the monopole, dipole and Gaussian form factor 
parametrizations with $r_p$ set to the muonic Lamb shift result of 0.84~fm.
Here the low-$Q^2$ points are the same as in Fig.~\ref{TheZoom},
the intermediate $Q^2$ data are from Mainz14 \cite{Bernauer:2013tpr} and
the highest $Q^2$ data are from Jefferson~Lab04 \cite{Christy:2004rc}
and Stanford94 \cite{Andivahis:1994rq}.  On this scale, the 0.84~fm 
dipole function is indistinguishable from the four-parameter 
continued-fraction fit of Griffioen, Carlson and Maddox
\cite{Griffioen:2015hta}.}
  \label{full}
\end{figure}

\begin{figure}[hbtp]
  \centering
  \includegraphics[width=\linewidth]{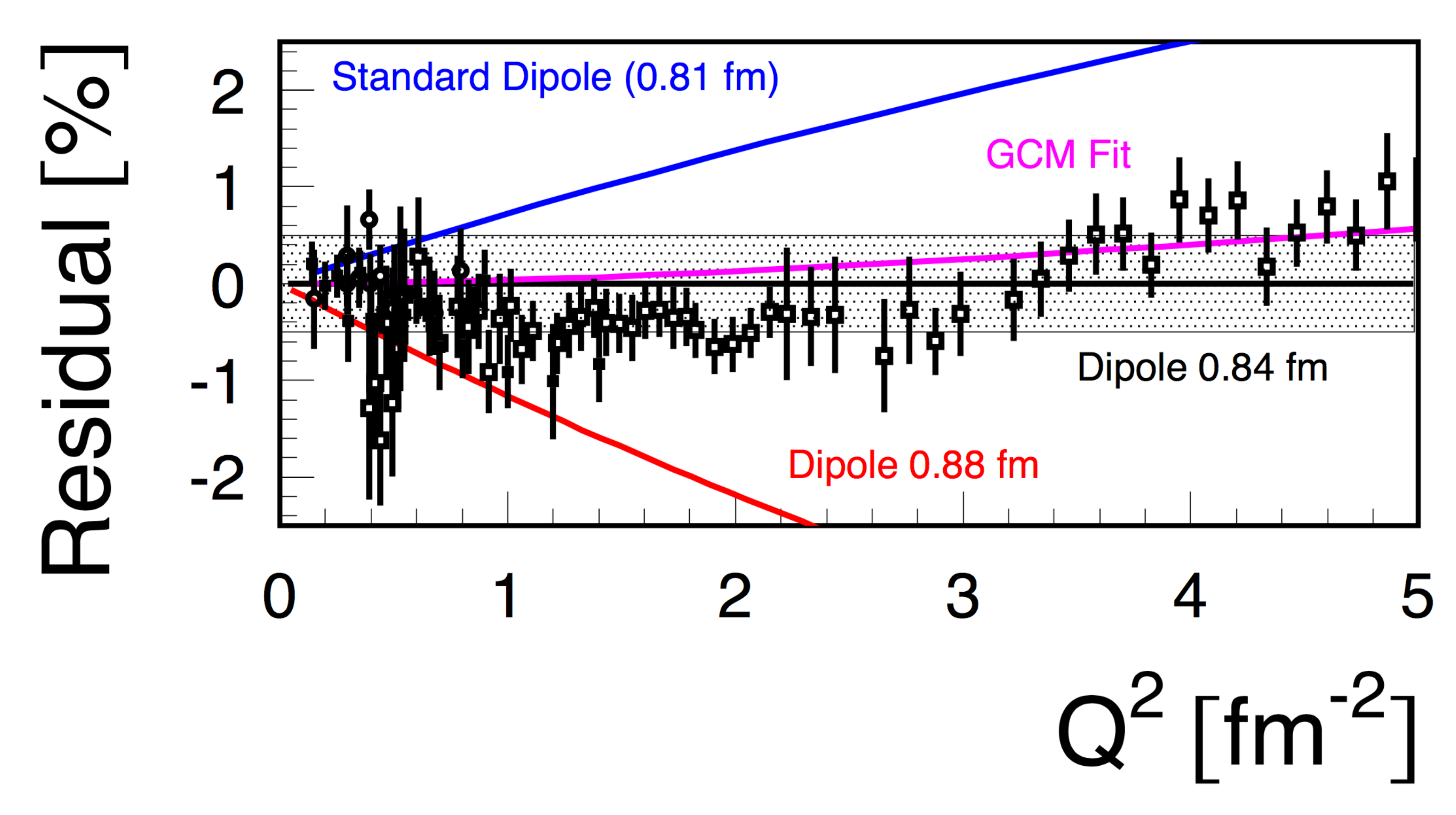}
  \caption{Shown are the residuals, $f(r_p)/f(0.84~{\mathrm{fm}})-1$, 
to the 0.84~fm dipole function for the published experimental values 
of $G_E(Q^2)$ along with various dipole forms.   Also shown 
is the residual to the GCM fit \cite{Griffioen:2015hta} with $G_E(0)= 1$.
The shading indicates a $\pm$~0.5\% band around the 0.84~fm dipole 
function to indicate the uncertainty of the experimental normalization, $n_0$.
While the classic 0.81~fm dipole function is ruled out, the data clearly 
follow the 0.84~fm dipole function.}
  \label{residual}
\end{figure}

In Fig.~\ref{full}, we plot $G_E$ for all the data we have studied 
along with the monopole, dipole and Gaussian functions
with $r_p$ set to 0.84~fm in each case.   The general agreement 
between the dipole function and the data is striking.
In Fig.~\ref{residual}, we zoom in on the low-$Q^2$ region and plot
the fractional difference between the data and the dipole function
corresponding to $r_p$ = 0.84~fm as well as between dipole functions
corresponding to the standard (CODATA) $r_p$ = 0.81~fm (0.88~fm) dipole
function.  Note that both the standard and the CODATA dipole functions 
fail to describe the data, while the $r_p = 0.84$~fm dipole 
function is within normalization uncertainties and does indeed adequately 
describe the data.  While agreement of the extracted radius, employed 
in the dipole function, for values of $Q^2$ beyond the low-$Q^2$ region is, 
strictly speaking, not germane to the determination of $r_p$ using 
statistical methods, it does lend support for our initial results 
and to our statistical approach to the analysis.

In summary, we have analyzed several high-quality measurements 
of the proton charge form factor using stepwise linear regression.
By performing numerous fits over several intervals of $Q^2$, 
we find that statistically justified linear extrapolations
of the extremely low-$Q^2$ data produce a proton charge radius 
which is consistent with the muonic results
and is systematically smaller than the one extracted using 
higher order extrapolation functions.
We also find that the simple dipole function incorporating 
the muonic radius provides a significantly better description
of the charge form factor than the often used standard dipole.   
While the uncertainties in the charge radius extracted from the data 
at finite $Q^2$ are not mathematically rigorously determined at $Q^2=0$, 
by using multivariate estimates of the uncertainties we tried to provide 
a reasonable error estimate.

\begin{appendix}

\section{Test of Adding Terms in a Polynomial Fit and Inspection of Results}
\label{polyapp}

A common problem when fitting data is determining the level 
of complexity of the fitting function required to accurately model 
the data. The $\chi^2$ statistic alone is not sufficient to determine 
if more complexity is required as $\chi^2$ will generally improve 
with increased complexity, i.e.~the addition of more parameters. 
In the main paper, the decision to apply a linear model to the fits 
of the Mainz80 and Saskatoon74 $G_E(Q^2)$ data over the $Q^2$ 
interval [0.14,0.8]~fm$^{-2}$ was founded on the observation that 
the data, when plotted against $Q^2$, appears linear and the assumption 
that higher order terms in a polynomial model may be neglected 
at low $Q^2$. Because general polynomials of degree $m$ and $m+1$ 
are nested, the significance of the higher degree polynomial 
may be tested using a statistical test such as the $F$-test.

In general, an $F$-test is simply any statistical test where the test statistic has an $F$-distribution 
under the null hypothesis~\cite{Behnke13,Mandel12}.
This is a common way of determining the statistical significance of additional complexity of the fitting function. In 
this case, we use the test since the true function, $G_E(Q^2)$, is not known and thus $\chi^2$ alone
is not sufficient for determining the degree one should use. 

To approximate the true function, $G_E(Q^2)$, we are using the Maclaurin series 
\begin{eqnarray*}
f(Q^2) &=& n_0  G_E(Q^2) \\
  &\approx & n_0 \left( 1 + a_1 x + a_2 x^2 + \cdots + a_m x^m \right) \>,
\end{eqnarray*}
where $m$ is the highest degree of the polynomial in $x=Q^2$ and $n_0$
is to account for normalization offsets.

\subsection{Low-$Q^2$ $F$-Test}

Theoretically, the radius of the proton can be determined by measuring 
the slope of $G_E(Q^2)$ at $Q^2=0$.  Thus, in the low $Q^2$ region, 
it is perhaps justified to only consider terms of order $Q^2$ 
and neglect higher order terms. To test this ansatz, we consider 
the two nested functions  
\begin{eqnarray}
f_1(Q^2) &=& n_0 \left( 1 + a_1 Q^2 \right) \>, \label{test1} \\ 
f_2(Q^2) &=& n_0 \left( 1 + a_1 Q^2 + a_2 Q^4 \right) \>, \label{test2}
\end{eqnarray}
where $n_0$ is simply a coefficient used to normalize the data 
such that at $Q^2=0$ one has $G_E(0) = 1$.  We are checking to see 
if the $a_2$ term should be included in the fit.

For the $F$-test, we use the ratio
\begin{equation}
F = \frac{{\chi^2}(j-1) - \chi^2(j)}{\chi^2(j)}  (N-j-1),
\end{equation}
where $N$ is the number of data points, $j$ the number of parameters 
being fit, and $\chi^2(j-1)$ and $\chi^2(j)$ are the total $\chi^2$ 
obtained from fitting at $j-1$ and $j$.  This test is applied to determine, 
to a specified confidence level, if the next order of a given function
should be used to describe the data or rejected.   
For the case of power series, the calculation of $F$ could even be 
built into a robust linear regression program, as suggested 
by Bevington \cite{Bevington03}, and the number of terms in the series 
determined automatically.   

In an $F$-test, if the simpler function is sufficient (which is the null 
hypothesis) it is expected that the relative increase in the $\chi^2$ 
is approximately proportional to the relative increase in the degrees 
of freedom.  The test may be applied when the functions under comparison 
are linear in the parameters ($a_i$) and when the simpler function 
can be nested in a function of more generality 
\cite{Bendat:2012,Mandel12,Behnke13}.  No stipulations 
on the correlations between the fit parameters are required.

When comparing the fit results between ($j=2$) and ($j=3$), 
as shown in Table~\ref{lowQFtest}, we obtain $F=0.00025$.   
This is far less than the $F$-distribution $\mathrm{CL}=95\%$ 
critical value of 4.3 (see Ref.~\cite{James:2006}, Table 10.2 
or Ref.~\cite{Bevington03}, Table C.5) and the more complicated 
function (Eq.~\ref{test2}) is rejected.

\begin{table}
\caption{Comparison of the nested functions with $j=2,3$ for 
a cut-off of 0.8~fm$^{-2}$.  The $j=3$ function is rejected 
at a $95\%$ confidence level by the $F$-test.
Smaller cut-offs, of course, also all reject the $j=3$ term.
Here $\nu = N-j-1$ is the number of degrees of freedom.   
For the uncertainty, we have used $\Delta\chi^2$ of 2.3 and 3.5 
for the 2- and 3-parameters fits, respectively,
in order to maintain a 68.3\% probability content.}
\label{lowQFtest}
\begin{ruledtabular}
\begin{tabular}{ccccccc}
$N$&$j$& $\chi^2$  & $\chi^2/{\nu}$ & $n_0$  & $a_1$        & $a_2$   \\
\hline
24 & 2 & 13.71     & 0.623          & 1.002(2) & $-0.119(4)$ &     \\
24 & 3 & 13.71     & 0.652          & 1.002(5) & $-0.120(20)$  & 0.00(2) \\
\end{tabular}
\end{ruledtabular}
\end{table}

In the analysis by Griffioen, Carlson and Maddox, the authors fit 
the low-$Q^2$ Mainz14 data with a power series.    
In their linear vs.~quadratic fits, they show very little change 
in $\chi^2/{\nu}$ for the extra degree of freedom.  In fact, 
applying the $F$-test to their results once again suggests that 
within this range the $a_2$ term is not needed.   

\subsection{Moderate-$Q^2$ $F$-Test}

Of course, as the range in $Q^2$ is increased, the higher order terms 
may be needed to fit the data. 
As an example, we consider the Mainz14 Rosenbluth $G_E$ data.  In this case, 
the data range from 0.39 to 14.2 fm$^{-2}$ and likely require a relatively high 
order Maclaurin series to be fitted precisely.

\begin{table*}
\caption{Comparison of the nested Maclaurin functions $j=5,6,7$ for 
the Mainz14 Rosenbluth $G_E$ data, 0.39 to 14.2 fm$^{-2}$, where 
a 0.2\% systematic uncertainty has been included.   Using the $F$-test, 
the $j=7$ function is rejected.    The lower order fits should be inspected.
The 68.3\% uncertainties for the $j=5,6,7$ fits are calculated using 
$\Delta\chi^2=5.9,7.0,8.2$, respectively.}
\label{intermed}
\begin{ruledtabular}
\begin{tabular}{cccccccccccc}
$N$  & $j$  & $\chi^2$  & $\chi^2/{\nu}$ & $n_0$    &$a_1$       & $a_2$                  &$a_3$                   & $a_4$     & $a_5$     & $a_6$        \\
\hline
77 & 5  & 49.57     & 0.688          & 0.991(2) &$-0.113(1)$ & $0.88(1)\cdot 10^{-2}$ &$-0.44(2)\cdot 10^{-3}$ & $9.7(8) \cdot 10^{-6}$ &                          &                      \\ 
77 & 6  & 41.34     & 0.582          & 0.996(2) &$-0.121(1)$ & $1.25(1)\cdot 10^{-2}$ &$-1.14(2)\cdot 10^{-3}$ & $6.8(1) \cdot 10^{-5}$ & $-1.62(7) \cdot 10^{-6}$ &                      \\
77 & 7  & 41.32     & 0.590          & 0.995(3) &$-0.119(1)$ & $1.18(1)\cdot 10^{-2}$ &$-0.93(2)\cdot 10^{-3}$ & $3.9(1) \cdot 10^{-5}$ & $~0.12(6) \cdot 10^{-6}$ & $-4.2(5) \cdot 10^{-8}$ \\
\end{tabular}
\end{ruledtabular}
\end{table*}

Using a Maclaurin expansion of the standard dipole 
for the initialization parameters of the fit, we
find the results summarized in Table~\ref{intermed}.
In this case,  we reject the $j=7$ function, for which $F=0.02$.  
Note that the test is only for rejecting unneeded complexity and 
is not an acceptance test of which of the lower order functions 
should be used to describe the data and thus it is not correct
to simply select the highest order function that is not rejected.

For a term-by-term comparison, we now write the 0.84~fm radius 
dipole function in terms of a Maclaurin expansion (note that
$(\hbar c)^2/0.66\,\mathrm{GeV}^2 \approx 0.0588\,\mathrm{fm}^2$):
\begin{eqnarray*}
\frac{1}{(1+0.0588 \, x)^{2}}  
  &=& 1 - 0.118 x + 1.04\!\cdot\!10^{-2} x^2 \\[-7pt]
  &-& 0.81 \!\cdot\! 10^{-3} x^3 + 5.98 \!\cdot\! 10^{-5} x^4 \\
  &-& 4.22 \!\cdot\! 10^{-6} x^5 
      + 2.89 \!\cdot\! 10^{-7} x^6 + \mathcal{O}\left( x^7 \right) \>.
\end{eqnarray*}
Comparing the coefficients of the Maclaurin expansion of the dipole 
function ($r_p\ =\ 0.84$~fm) with the corresponding fitted parameters, 
$a_n$, for the $j=5$ and $j=6$ cases (Table~\ref{intermed}),
one sees that the coefficients of the dipole function fall between 
those of the two fits to all overlapping orders.  This arises because 
the first neglected terms in the two cases are of opposite sign. 
Thus, in each case the remaining terms must adjust in opposite directions
to account for the neglected contribution, suggesting strongly that the
dipole coefficients are very close to the proper ones.

\section{ANOVA Tables and AIC Tests}
\label{ANOVA-tables}

There are certainly many other techniques that can be used.   
For example, it is common in statistics to generate an Analysis 
of Variance, ANOVA, table.  It is convenient to use  R~\cite{R} 
to generate the table, the ratio of the sum of squares, $F$, as well as 
to calculate the corresponding $p$-value (the probability 
of $F$ being higher than the critical value)
on which the significance code is based: smaller values of $p$
imply high significance and values near unity lower significance;
see Tables~\ref{ANOVA1}, \ref{ANOVA2} and \ref{ANOVA3}. 
Following R's standard notation, three stars signify a significant 
effect while no stars imply that the hypothesis can be rejected 
with a high degree of certainty.

While using a slightly different criteria, these tables support
the same conclusions as we draw from our $F$-test, namely, that 
for the very low-$Q^2$ data the only statistical model that is justified 
is the linear one.  This result neither confirms nor invalidates 
any physical models, as all the physical models possess a linear term
in their Maclaurin expansions.  The difference is that the statistical model 
is only valid within the range of the data while a physical model can have 
a significance over all $Q^2$.  

What one can see is that, as one increases the range of the data, 
one can add terms to the statistical model.   While this makes perfect 
sense for describing the range of the data, it also means that 
as the range of the data increases, higher and higher order functions 
are being used to find the slope and intercept to $Q^2 = 0$ by extrapolation.

\begin{table}[hbtp]
\caption{Standard ANOVA table generated by R for the linear-regression
models fits of the limited (0.14 to 0.8~fm$^{-2}$) Saskatoon74 and Mainz80 
data.    Here there is no question that the data only justify being fit 
with the linear statistical model.}
\label{ANOVA1}
\begin{ruledtabular}
\begin{tabular}{ccccccc}
$j$ & DOF &    RSS &  Sum of Sq &     $F$  & $p$-value &   \\
\hline
2 &    22  & 13.706 &           &        &          &   \\
3 &    21 &  13.706 &  0.00017  & 0.0003 & 0.9875   & $-$ \\
\end{tabular}
\end{ruledtabular}
\end{table}

\begin{table}[hbtp]
\caption{Standard ANOVA table generated by R for the linear-regression
model fits of the full Saskatoon74 and Mainz80 data (0.14 to 1.4~fm$^{-2}$).   
Here the two star significance of the polynomial justifies trying 
the polynomial function for that range as done within the paper.}
\label{ANOVA2}
\begin{ruledtabular}
\begin{tabular}{ccccccc}
$j$ &  DOF &   RSS  & Sum of Sq &      $F$ &   $p$-value &    \\  
\hline
2 &    27   & 24.169  &           &        &            &    \\
3 &    26   & 18.869  &  5.2995   & 7.4894 & 0.01236    & **  \\
4 &    25   & 18.835  &  0.0344   & 0.0486 & 0.82768    & $-$  \\
\end{tabular}
\end{ruledtabular}
\end{table}

\begin{table}[btp]
\caption{Standard ANOVA table generated by R for the linear-regression
model fits of the complete Mainz14 data, with significance codes based 
on $p$-values.}
\label{ANOVA3}
\begin{ruledtabular}
\begin{tabular}{cccccccc}
$j$ &  DOF &   RSS & Sum of Sq &          $F$  &   $p$-value  &     \\
\hline
2 &   75    & 45842 &            &             &             &     \\       
3 &   74    &  1838 &    44004   & 19570.6  & $< 2.2\cdot 10^{-16}$ & *** \\
4 &   73    &   289 &     1549   &   688.9  & $< 2.2\cdot 10^{-16}$ & *** \\
5 &   72    &   191 &       98   &    43.7  & $6.623\cdot 10^{-9}$ & *** \\
6 &   71    &   159 &       32   &    14.1641  & 0.0003482 & *** \\
7 &   70    &   159 & $\approx 0$ &     0.0095  & 0.9224837 & $-$   \\
\end{tabular}
\end{ruledtabular}
\end{table}

\begin{figure}
\includegraphics[width=\linewidth]{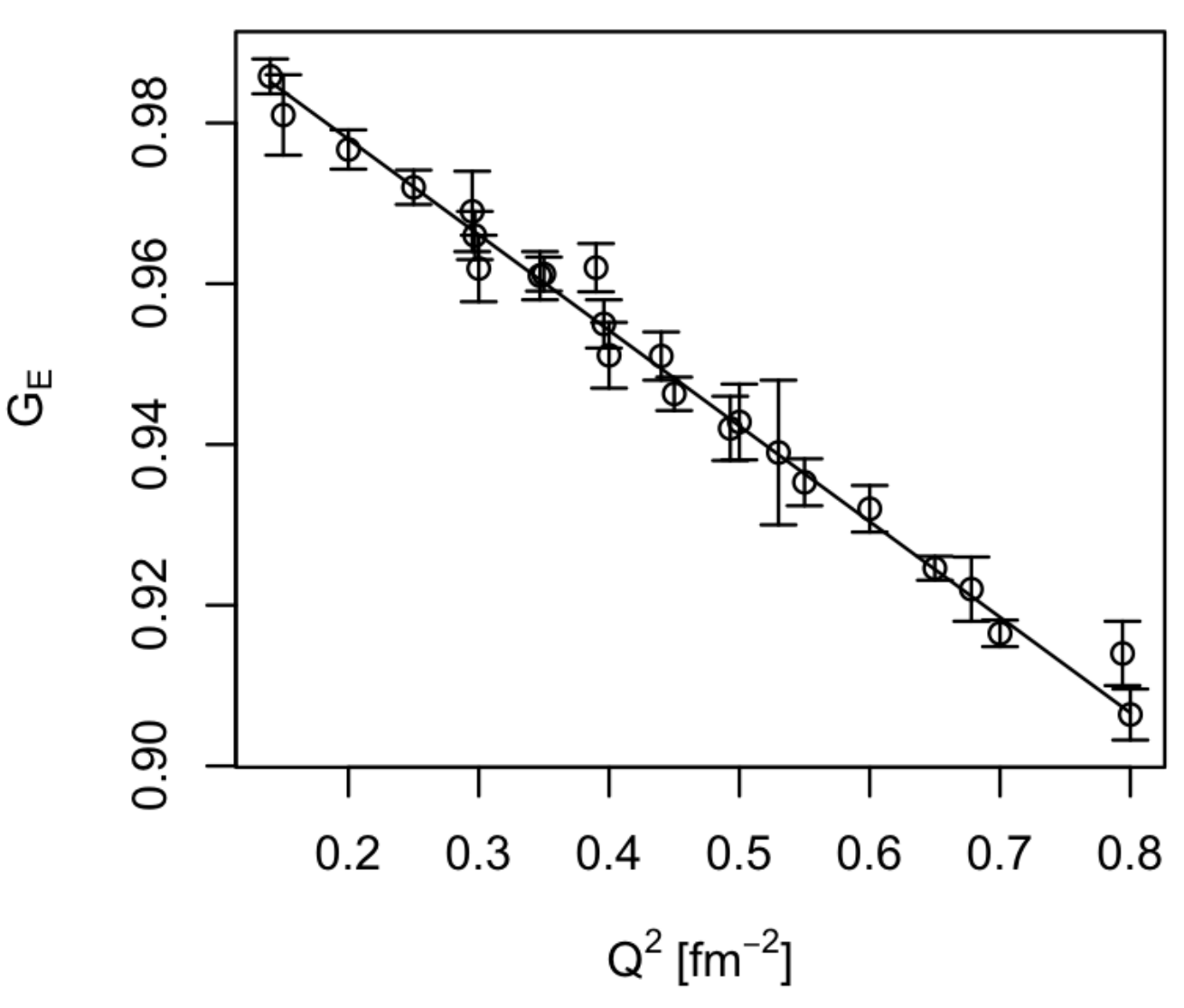}
\includegraphics[width=\linewidth]{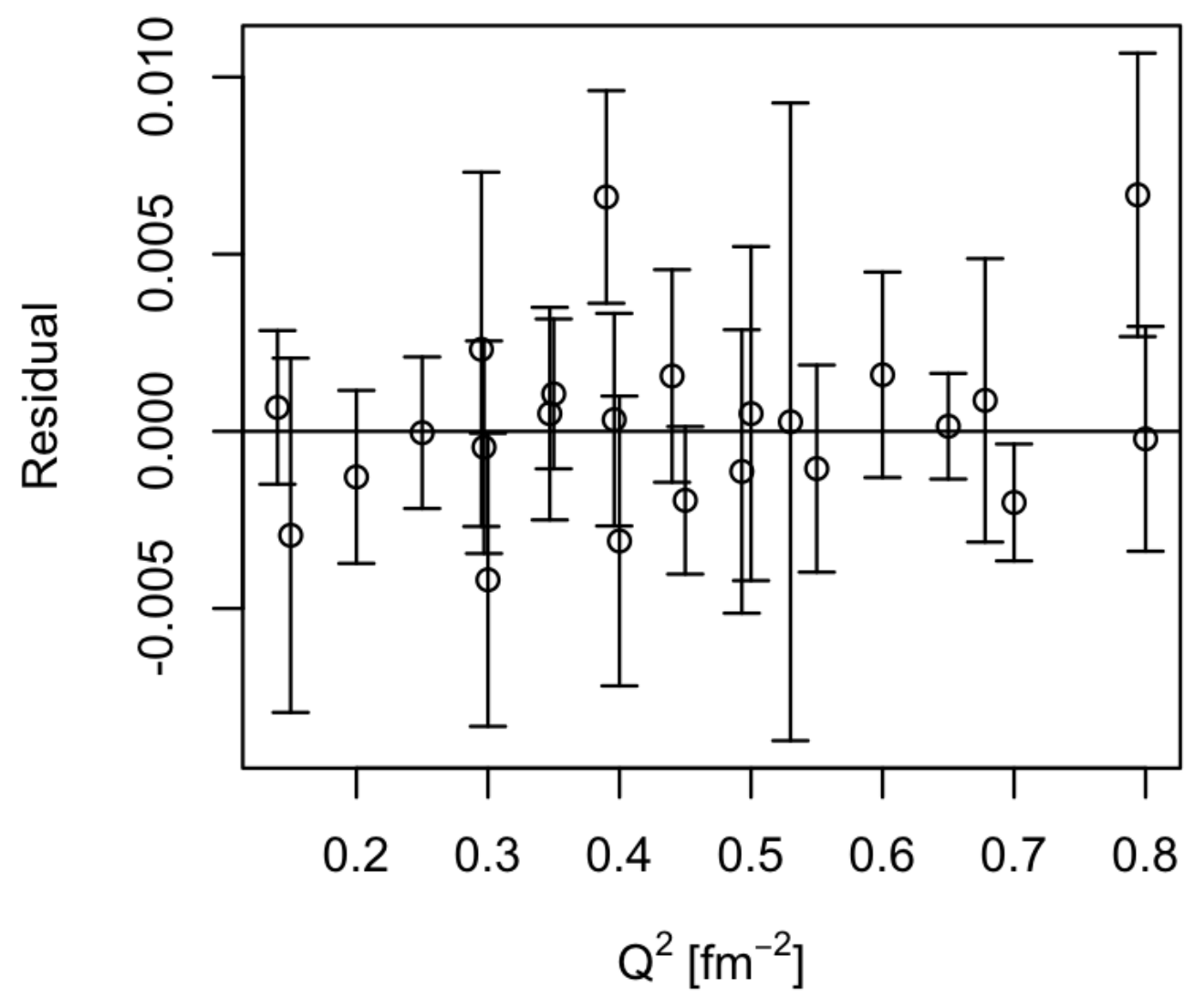}
\caption{Shown is the result of stepwise regression of the combined 
and truncated Saskatoon74 and Mainz80 data using AIC to determine 
that a degree-one polynomial sufficiently describes the data.
This log-likelihood based result is in agreement with the $\chi^2$-based 
$F$-test.}
\label{lowQ2-aic}
\end{figure}

\begin{figure}
\includegraphics[width=\linewidth]{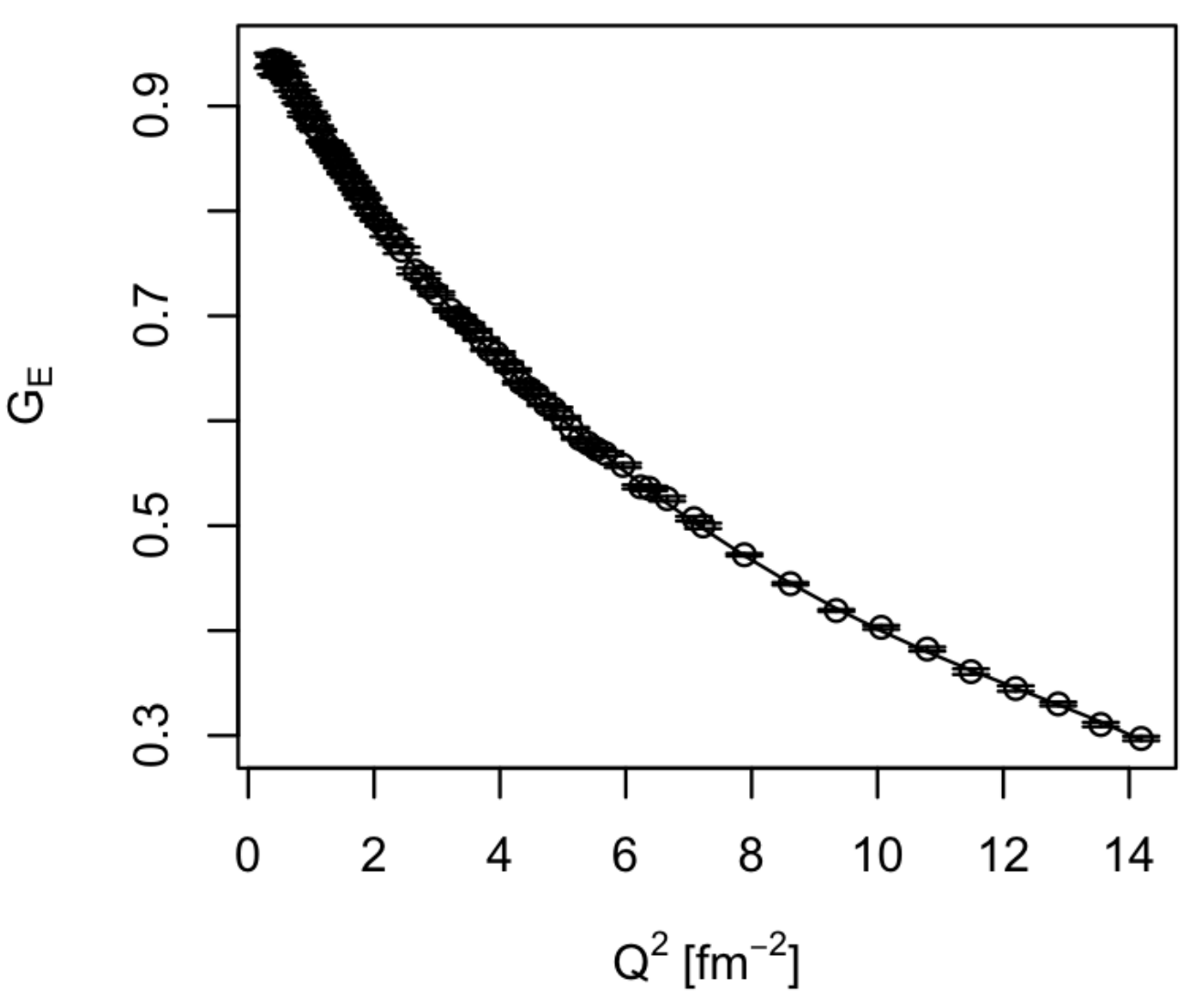}
\includegraphics[width=\linewidth]{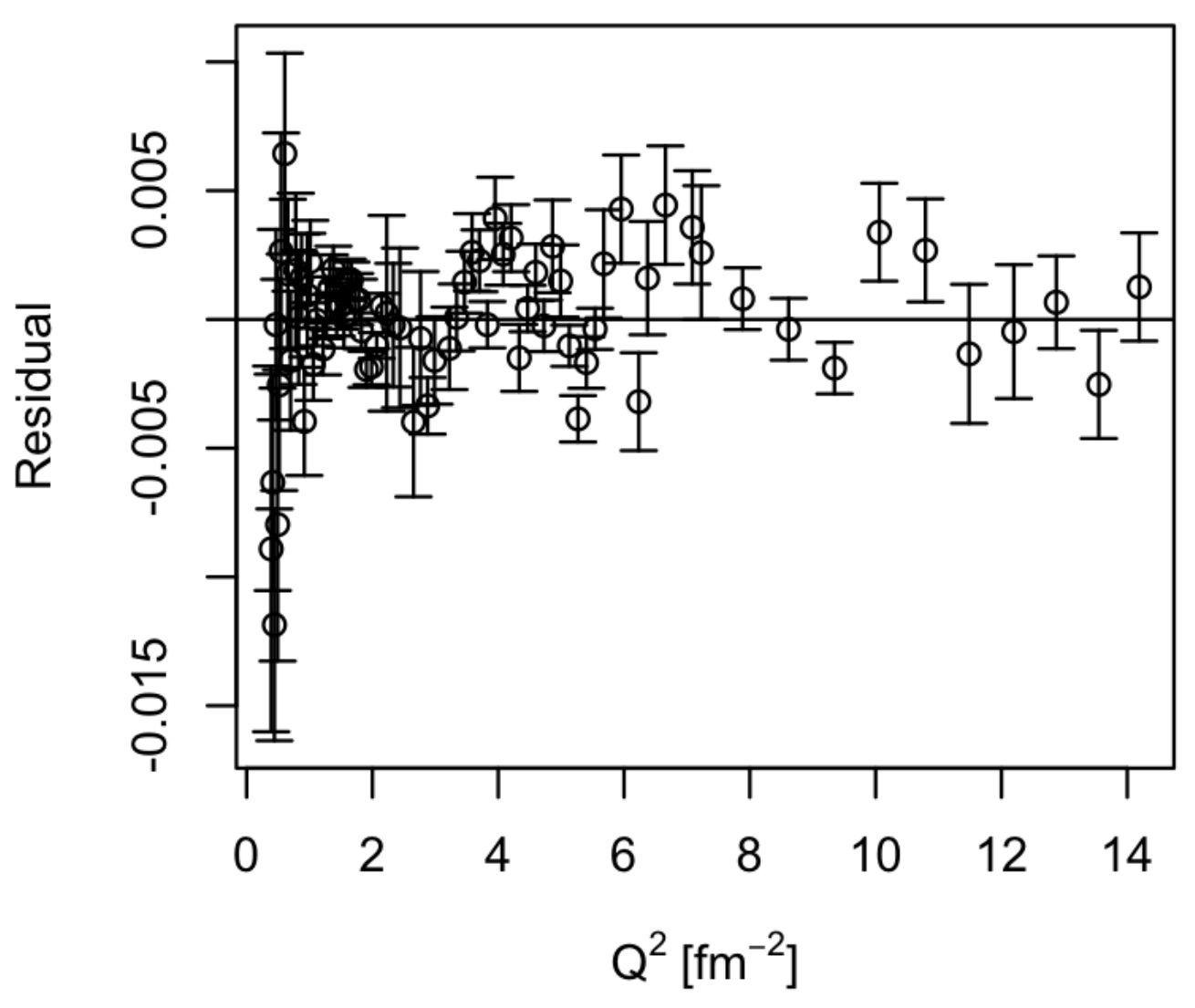}
\caption{Shown is the result of stepwise regression of the Mainz14 data 
using AIC to determine that a $5^\mathrm{th}$ degree polynomial 
sufficiently describes the data.  This log-likelihood based result
is in agreement with the $\chi^2$-based $F$-test.}
\label{mainz14-aic}
\end{figure}

Going beyond just the ANOVA tables, it is possible to do the stepwise 
linear regression using the Akaike information criterion (AIC).   
The beauty of using AIC is that non-nested models can be compared though
for the following two examples the models are the nested Maclaurin series.   
In Fig.~\ref{lowQ2-aic} the results of a stepwise linear
regression of the Saskatoon74 and Mainz80 data is presented
and, using the same code, Fig.~\ref{mainz14-aic} shows a stepwise
linear regression of the Mainz14 $G_E$ data. 
For both of these fits, the order of polynomial to use was chosen
using AIC.
The code to produce these results as well as to make the ANOVA tables  
makes use of the R programming language for statistical computing~\cite{R} 
and is freely available on GitHub \cite{Higinbotham:2016}.

\section{Multivariate Errors}

By default, MINUIT's MINOS algorithm performs minimization in multiple 
dimensions; however, the errors which it calculates by default are still 
only single-parameter errors.   In other words, the one-sigma uncertainty 
when the other parameters are allowed to vary to infinity.    Now if 
the intention of the fit is to make simultaneous statements about 
multiple parameters, the error estimate is complicated by the fact 
that the confidence region is not simply an interval on an axis but 
rather a hyper-volume.   Thus, it left to the user to set the desired 
probability content of the hyper-volume and the code will calculate 
the errors that correspond to such a content.  This prescription 
for handling multivariate errors is in fact explicitly noted as required 
in the statistics section of the Review of Particle Physics~\cite{PDG:2014}.

For the proton radius experiments, even the linear fits are trying 
to simultaneously determine the slope and normalization 
at $Q^2=0$ and thus we required a $\Delta\chi^2$ of 2.3 instead 
of the default 1 in order to keep the desired 68\% probability content 
inside the probability contour.  Further details and examples of this 
type of multivariate error analysis can be found in Y.~Avni's 
highly cited article on the analysis of x-ray spectra 
in galactic clusters \cite{Avni:1976}.

\end{appendix}

\begin{acknowledgments}
We thank Jan Bernauer, Carl Carlson, Larry Cardman, Tim Gay, and
Simon \v{S}irca for many useful discussions and Dennis Skopik for 
answering our questions about the Saskatoon data.
The fits were performed with the MINUIT package~\cite{James:1975dr},
R statistical computing language~\cite{R}, and Wolfram's Mathematica.
We explicitly use 68.3\% multivariate probability contents 
throughout this work.  This material is based upon work supported 
by the U.S. Department of Energy, Office of Science, Office of Nuclear 
Physics under contract {DE-AC05-060R23177} and {DE-SC0014325}.
\end{acknowledgments}


\begin{thebibliography}{10}
\providecommand*{\bibinfo}[2]{#2}
\providecommand*{\eprint}[1]{#1}
\providecommand*{\url}[1]{#1}
\bibitem{Pohl:2010}
\bibinfo{author}{R.~Pohl} \emph{et~al.}, \bibinfo{journal}{Nature}
  \bibinfo{volume}{\textbf{466}}, \bibinfo{pages}{213} (\bibinfo{date}{2010}).
\bibitem{Antognini:2012}
\bibinfo{author}{A.~Antognini} \emph{et~al.}, \bibinfo{journal}{Science}
  \bibinfo{volume}{\textbf{339}}, \bibinfo{pages}{417} (\bibinfo{date}{2013}).
\bibitem{Mohr:2012}
\bibinfo{author}{P.~J. Mohr}, \bibinfo{author}{B.~N. Taylor}, and
  \bibinfo{author}{D.~B. Newell}, \bibinfo{journal}{Rev. Mod. Phys.}
  \bibinfo{volume}{\textbf{84}}, \bibinfo{pages}{1527} (\bibinfo{date}{2012}).
\bibitem{Bernauer:2013tpr}
\bibinfo{author}{J.~C. Bernauer} \emph{et~al.}, \bibinfo{journal}{Phys. Rev. C}
  \bibinfo{volume}{\textbf{90}}, \bibinfo{pages}{015206}
  (\bibinfo{date}{2014}).
\bibitem{Pohl:2013yb}
\bibinfo{author}{R.~Pohl}, \bibinfo{author}{R.~Gilman}, \bibinfo{author}{G.~A.
  Miller}, and \bibinfo{author}{K.~Pachucki}, \bibinfo{journal}{Ann. Rev. Nucl.
  Part. Sci.} \bibinfo{volume}{\textbf{63}}, \bibinfo{pages}{175}
  (\bibinfo{date}{2013}).
\bibitem{Carlson:2015jba}
\bibinfo{author}{C.~E. Carlson}, \bibinfo{journal}{Prog. Part. Nucl. Phys.}
  \bibinfo{volume}{\textbf{82}}, \bibinfo{pages}{59} (\bibinfo{date}{2015}).
\bibitem{Griffioen:2015hta}
\bibinfo{author}{K.~Griffioen}, \bibinfo{author}{C.~Carlson}, and
  \bibinfo{author}{S.~Maddox} (\bibinfo{date}{2015}),
  \eprint{arXiv:1509.06676}.
\bibitem{Horbatsch:2015qda}
\bibinfo{author}{M.~Horbatsch} and \bibinfo{author}{E.~A. Hessels},
  \bibinfo{journal}{Phys. Rev. C} \bibinfo{volume}{\textbf{93}}(1),
  \bibinfo{pages}{015204} (\bibinfo{date}{2016}).
\bibitem{Lee:2015jqa}
\bibinfo{author}{G.~Lee}, \bibinfo{author}{J.~R. Arrington}, and
  \bibinfo{author}{R.~J. Hill}, \bibinfo{journal}{Phys. Rev. D}
  \bibinfo{volume}{\textbf{92}}, \bibinfo{pages}{013013}
  (\bibinfo{date}{2015}).
\bibitem{Lorenz:2014yda}
\bibinfo{author}{I.~T. Lorenz}, \bibinfo{author}{U.-G. Mei{\ss}ner},
  \bibinfo{author}{H.~W. Hammer}, and \bibinfo{author}{Y.~B. Dong},
  \bibinfo{journal}{Phys. Rev. D} \bibinfo{volume}{\textbf{91}},
  \bibinfo{pages}{014023} (\bibinfo{date}{2015}).
\bibitem{Simon:1980hu}
\bibinfo{author}{G.~G. Simon}, \bibinfo{author}{C.~Schmitt},
  \bibinfo{author}{F.~Borkowski}, and \bibinfo{author}{V.~H. Walther},
  \bibinfo{journal}{Nucl. Phys. A} \bibinfo{volume}{\textbf{333}},
  \bibinfo{pages}{381} (\bibinfo{date}{1980}).
\bibitem{Murphy:1974zz}
\bibinfo{author}{J.~J. Murphy}, \bibinfo{author}{Y.~M. Shin}, and
  \bibinfo{author}{D.~M. Skopik}, \bibinfo{journal}{Phys. Rev. C}
  \bibinfo{volume}{\textbf{9}}, \bibinfo{pages}{2125} (\bibinfo{date}{1974}),
  [Erratum: Phys. Rev. C {\bf 10}, 2111 (1974)].
\bibitem{Hand:1963zz}
\bibinfo{author}{L.~N. Hand}, \bibinfo{author}{D.~G. Miller}, and
  \bibinfo{author}{R.~Wilson}, \bibinfo{journal}{Rev. Mod. Phys.}
  \bibinfo{volume}{\textbf{35}}, \bibinfo{pages}{335} (\bibinfo{date}{1963}).
\bibitem{Arrington:2003qk}
\bibinfo{author}{J.~Arrington}, \bibinfo{journal}{Phys. Rev. C}
  \bibinfo{volume}{\textbf{69}}, \bibinfo{pages}{022201}
  (\bibinfo{date}{2004}).
\bibitem{Arrington:2007ux}
\bibinfo{author}{J.~Arrington}, \bibinfo{author}{W.~Melnitchouk}, and
  \bibinfo{author}{J.~A. Tjon}, \bibinfo{journal}{Phys. Rev. C}
  \bibinfo{volume}{\textbf{76}}, \bibinfo{pages}{035205}
  (\bibinfo{date}{2007}).
\bibitem{DeJager:1987qc}
\bibinfo{author}{H.~De~Vries}, \bibinfo{author}{C.~W. De~Jager}, and
  \bibinfo{author}{C.~De~Vries}, \bibinfo{journal}{Atom. Data Nucl. Data Tabl.}
  \bibinfo{volume}{\textbf{36}}, \bibinfo{pages}{495} (\bibinfo{date}{1987}).
\bibitem{Cardman:1980}
\bibinfo{author}{L.~S. Cardman} \emph{et~al.}, \bibinfo{journal}{Phys. Lett. B}
  \bibinfo{volume}{\textbf{91}}, \bibinfo{pages}{203} (\bibinfo{date}{1980}).
\bibitem{Miller:2007uy}
\bibinfo{author}{G.~A. Miller}, \bibinfo{journal}{Phys. Rev. Lett.}
  \bibinfo{volume}{\textbf{99}}, \bibinfo{pages}{112001}
  (\bibinfo{date}{2007}).
\bibitem{Kraus:2014qua}
\bibinfo{author}{E.~Kraus}, \bibinfo{author}{K.~E. Mesick},
  \bibinfo{author}{A.~White}, \bibinfo{author}{R.~Gilman}, and
  \bibinfo{author}{S.~Strauch}, \bibinfo{journal}{Phys. Rev. C}
  \bibinfo{volume}{\textbf{90}}, \bibinfo{pages}{045206}
  (\bibinfo{date}{2014}).
\bibitem{Higinbotham:2016}
\bibinfo{author}{D.~W. Higinbotham}, \bibinfo{title}{\emph{Model selection with
  stepwise regression}}, \url{http://jeffersonlab.github.io/model-selection/}
  (\bibinfo{date}{2016}).
\bibitem{James:2006}
\bibinfo{author}{F.~James}, \bibinfo{title}{\emph{{S}tatistical {M}ethods in
  {E}xperimental {P}hysics}} (\bibinfo{publisher}{World Scientific},
  \bibinfo{year}{2006}), 2nd Edition.
\bibitem{Bernauer:2010wm}
\bibinfo{author}{J.~C. Bernauer} \emph{et~al.} (\bibinfo{collaboration}{A1}),
  \bibinfo{journal}{Phys. Rev. Lett.} \bibinfo{volume}{\textbf{105}},
  \bibinfo{pages}{242001} (\bibinfo{date}{2010}).
\bibitem{PDG:2014}
\bibinfo{author}{K.~Olive} \emph{et~al.} (\bibinfo{collaboration}{Particle Data
  Group}), \bibinfo{journal}{Chin. Phys. C} \bibinfo{volume}{\textbf{38}},
  \bibinfo{pages}{090001} (\bibinfo{date}{2014}).
\bibitem{sirca:2012}
\bibinfo{author}{S.~\v{S}irca} and \bibinfo{author}{M.~Horvat},
  \bibinfo{title}{\emph{Computational Methods for Physicists}}
  (\bibinfo{publisher}{Springer}, \bibinfo{year}{2012}).
\bibitem{Puckett:2010ac}
\bibinfo{author}{A.~J.~R. Puckett} \emph{et~al.}, \bibinfo{journal}{Phys. Rev.
  Lett.} \bibinfo{volume}{\textbf{104}}, \bibinfo{pages}{242301}
  (\bibinfo{date}{2010}).
\bibitem{Andivahis:1994rq}
\bibinfo{author}{L.~Andivahis} \emph{et~al.}, \bibinfo{journal}{Phys. Rev. D}
  \bibinfo{volume}{\textbf{50}}, \bibinfo{pages}{5491} (\bibinfo{date}{1994}).
\bibitem{Christy:2004rc}
\bibinfo{author}{M.~E. Christy} \emph{et~al.}, \bibinfo{journal}{Phys. Rev. C}
  \bibinfo{volume}{\textbf{70}}, \bibinfo{pages}{015206}
  (\bibinfo{date}{2004}).
\bibitem{Note1}
While a large reduced $\chi ^2$ can be used to reject a model with a high
  degree of certainty, $\chi ^2$ alone should not be used for choosing between
  models that have not been rejected.
\bibitem{Behnke13}
\bibinfo{author}{G.~S. O.~Behnke, K.~Kroninger} and
  \bibinfo{author}{T.~Schomer-Sadenius}, \bibinfo{title}{\emph{Data Analysis in
  High Energy Physics: A Practical Guide to Statistical Methods}}
  (\bibinfo{publisher}{Wiley}, \bibinfo{year}{2013}).
\bibitem{Mandel12}
\bibinfo{author}{J.~Mandel}, \bibinfo{title}{\emph{The Statistical Analysis of
  Experimental Data}} (\bibinfo{publisher}{Courier Corporation},
  \bibinfo{year}{2012}).
\bibitem{Bevington03}
\bibinfo{author}{P.~R. Bevington} and \bibinfo{author}{D.~Robinson},
  \bibinfo{title}{\emph{{D}ata {R}eduction and {E}rror {A}nalysis for the
  {P}hysical {S}ciences}} (\bibinfo{publisher}{McGraw-Hill},
  \bibinfo{year}{2003}), 3rd Edition.
\bibitem{Bendat:2012}
\bibinfo{author}{J.~Bendat} and \bibinfo{author}{A.~Piersol},
  \bibinfo{title}{\emph{Random Data: Analysis and Measurement Procedures}}
  (\bibinfo{publisher}{Wiley}, \bibinfo{year}{2012}).
\bibitem{R}
\bibinfo{author}{{R Core Team}}, \bibinfo{title}{\emph{R: A Language and
  Environment for Statistical Computing}}, R Foundation for Statistical
  Computing, Vienna, Austria (\bibinfo{date}{2015}),
  \url{https://www.R-project.org/}.
\bibitem{Avni:1976}
\bibinfo{author}{Y.~{Avni}}, \bibinfo{journal}{\apj}
  \bibinfo{volume}{\textbf{210}}, \bibinfo{pages}{642} (\bibinfo{date}{1976}).
\bibitem{James:1975dr}
\bibinfo{author}{F.~James} and \bibinfo{author}{M.~Roos},
  \bibinfo{journal}{Comput. Phys. Commun.} \bibinfo{volume}{\textbf{10}},
  \bibinfo{pages}{343} (\bibinfo{date}{1975}).

\end{thebibliography}

\end{document}